\documentclass[a4paper,journal,twocolumn,10pt,final]{IEEEtran}

\usepackage[none]{hyphenat}
\usepackage[utf8]{inputenc}
\usepackage[T1]{fontenc}
\usepackage{textcomp}
\usepackage{scalefnt}
\usepackage{amsmath,amsfonts,amssymb,amsbsy,amstext}

\let\originalleft\left
\let\originalright\right
\renewcommand{\left}{\mathopen{}\mathclose\bgroup\originalleft}
\renewcommand{\right}{\aftergroup\egroup\originalright}

\usepackage{graphicx}
\graphicspath{{.}{.\figs}}

\usepackage{multirow}
\usepackage{tabularx}
\usepackage{booktabs}
\newcolumntype{C}{>{\centering\arraybackslash}X}
\newcolumntype{R}{>{\flushright\arraybackslash}X}
\newcolumntype{L}{>{\flushleft\arraybackslash}X}
\newcolumntype{P}{>{\centering\arraybackslash} p{0.5\linewidth}}
\usepackage{mathtools, cuted}
\newcommand{\mat}[1]{\mbox{\boldmath{$#1$}}}
\usepackage[%
style=ieee,
backend=biber, %
]{biblatex}
\defbibheading{bibliography}[\refname]{\section*{#1}}

\makeatletter
\g@addto@macro{\UrlBreaks}{\UrlOrds}
\makeatother

\usepackage{acro}

\NewDocumentCommand{\acro}{m o m o}
{%
	\IfValueTF{#2}{%
		\IfValueTF{#4}{%
			\DeclareAcronym{#1}{short={#2},long={#3},#4}
		}{%
			\DeclareAcronym{#1}{short={#2},long={#3}}
		}
	}{%
		\IfValueTF{#4}{%
			\DeclareAcronym{#1}{short={#1},long={#3},#4}
		}{%
			\DeclareAcronym{#1}{short={#1},long={#3}}
		}
	}
}

\acro{1G}{First Generation}
\acro{2G}{Second Generation}
\acro{3G}{Third Generation}
\acro{4G}{Fourth Generation}
\acro{5G}{Fifth Generation}
\acro{5GC}{5G Core Network}
\acro{3GPP}{3rd Generation Partnership Project}
\acro{3GPP2}{3rd Generation Partnership Project 2}
\acro{AA}{Antenna Array}
\acro{AC}{Admission Control}
\acro{AD}{Attack-Decay}
\acro{AF}{Amplify and Forward}
\acro{ABS}{Almost Blank Subframe}
\acro{ADSL}{Asymmetric Digital Subscriber Line}
\acro{AHW}{Alternate Hop-and-Wait}
\acro{AI}{Artificial Intelligence}
\acro{AMC}{Adaptive Modulation and Coding}
\acro{AP}{Access Point}
\acro{APA}{Adaptive Power Allocation}
\acro{ARMA}{Autoregressive Moving Average}
\acro{ASC}{Adaptive Satisfaction Control}
\acro{ATES}{Adaptive Throughput-based Efficiency-Satisfaction Trade-Off}
\acro{AWGN}{Additive White Gaussian Noise}
\acro{BB}{Branch and Bound}
\acro{BC}{Branch and Cut}
\acro{BD}{Block Diagonalization}
\acro{BER}{Bit Error Rate}
\acro{DNN}{Deep Neural Network}
\acro{BF}{Best Fit}
\acro{BL}{Bit Loading}
\acro{BLER}{BLock Error Rate}
\acro{BLPC-1}{Bit Loading with Power Constraint in Hop 1}
\acro{BLPC-2}{Bit Loading with Power Constraint in Hop 2}
\acro{BPC}{Binary Power Control}
\acro{BPSK}{Binary Phase-Shift Keying}
\acro{BRA}{Balanced Random Allocation}
\acro{BS}{Base Station}
\acro{BSP}{\acs*{BS} Placement}
\acro{CAP}{Combinatorial Allocation Problem}
\acro{CAPEX}{Capital Expenditure}
\acro{CB}{Contextual Bandit}
\acro{CBF}{Coordinated Beamforming}
\acro{CBR}{Constant Bit Rate}
\acro{CBS}{Class Based Scheduling}
\acro{CC}{Congestion Control}
\acro{CCL}{Common Cell List}
\acro{CDF}{Cumulative Distribution Function}
\acro{CDMA}{Code-Division Multiple Access}
\acro{CH}{Channel Hardening}
\acro{CHO}{Conditional Handover}
\acro{C-RAN}{cloud-based Radio Access Network}
\acro{CL}{Closed Loop}
\acro{CLPC}{Closed Loop Power Control} 
\acro{CN}{Core Network}       
\acro{CNR}{Channel-to-Noise Ratio}
\acro{CPA}{Cellular Protection Algorithm}
\acro{CPICH}{Common Pilot Channel}
\acro{CoMP}{Coordinated Multi-Point}
\acro{CQI}{Channel Quality Indicator}
\acro{CRE}{Cell Range Expansion}
\acro{CRM}{Constrained Rate Maximization}
\acro{CRN}{Cognitive Radio Network}
\acro{C-RNTI}{Cell Radio Network Temporary Identifier}
\acro{CRRM}{Centralized/Common Radio Resource Management}
\acro{CS}{Coordinated Scheduling}
\acro{CSI}{Channel State Information}
\acro{CTS}{Clear to Send}
\acro{CUE}{Cellular User Equipment}
\acro{CWND}{Congestion window size}
\acro{D2D}{Device-to-Device}
\acro{DC}{Dual Connectivity}
\acro{DCA}{Dynamic Channel Allocation}
\acro{DE}{Differential Evolution}
\acro{DF}{Decode and Forward}
\acro{DFT}{Discrete Fourier Transform}
\acro{DIST}{Distance}
\acro{DL}{Downlink}
\acro{DMA}{Double Moving Average}
\acro{DMRS}{Demodulation Reference Signal}
\acro{D2DM}{\acs*{D2D} Mode}
\acro{DMS}{\acs*{D2D} Mode Selection}
\acro{DPC}{Dirty Paper Coding}
\acro{DQN}{Deep $Q$-Network}
\acro{DRA}{Dynamic Resource Assignment}
\acro{DRL}{Deep Reinforcement Learning}
\acro{DSA}{Dynamic Spectrum Access}
\acro{DSM}{Delay-based Satisfaction Maximization}
\acro{E2E}{End-to-End}
\acro{ECC}{Electronic Communications Committee}
\acro{EDF}{Earliest Deadline First}
\acro{EE}{Energy Efficiency}
\acro{EFLC}{Error Feedback Based Load Control}
\acro{EI}{Efficiency Indicator}
\acro{e-ICIC}{Enhanced Inter-Cell Interference Coordination}
\acro{eMBB}{Enhanced Mobile Broadband}
\acro{eNB}{Evolved Node B}
\acro{EXP}{Exponential}
\acro{EPA}{Equal Power Allocation}
\acro{EPC}{Evolved Packet Core}
\acro{EPS}{Evolved Packet System}
\acro{E-UTRAN}{Evolved Universal Terrestrial Radio Access Network}
\acro{ES}{Exhaustive Search}
\acro{FCP}{Fundamental Counting Principle}
\acro{FCA}{Flow Control Algorithm}
\acro{FD}{Full-Duplex Communications}
\acro{FDD}{Frequency Division Duplex}
\acro{FDM}{Frequency Division Multiplexing}
\acro{FDMA}{Frequency Division Multiple Access}
\acro{FER}{Frame Erasure Rate}
\acro{FIFO}{First In First Out}
\acro{FF}{Fast Fading}
\acro{FRS}[FS]{Fast-RAT Scheduling}
\acro{FS}{Fast Switching}
\acro{FSB}{Fixed Switched Beamforming}
\acro{FST}{Fixed \acs*{SNR} Target}
\acro{FTP}{File Transfer Protocol}
\acro{GA}{Genetic Algorithm}
\acro{GAP}{Generalized Assignment Problem}
\acro{GAP-MQ}{Generalized Assignment Problem with Minimum Quantities}
\acro{GATES}{Generalized Adaptive Throughput-based Efficiency-Satisfaction Trade-Off}
\acro{GBR}{Guaranteed Bit Rate}
\acro{GLR}{Gain to Leakage Ratio}
\acro{gNB}{gNode B}
\acro{GOS}{Generated Orthogonal Sequence}
\acro{GPL}{GNU General Public License}
\acro{GPS}{Global Positioning System}
\acro{GRP}{Grouping}
\acro{GSM}{Global System for Mobile Communications}
\acro{GTEL}{Wireless Telecommunications Research Group}
\acro{HARQ}{Hybrid Automatic Repeat Request}
\acro{HCPP}{Hardcore Point Process}
\acro{HD}{High Definition}
\acro{HetNet}{Heterogeneous Network}
\acro{HH}{Hughes-Hartogs}
\acro{HardH}[HH]{Hard Handover}
\acro{HMS}{Harmonic Mode Selection}
\acro{HO}{Handover}
\acro{HOL}{Head Of Line}
\acro{HPBW}{Half Power Beamwidth}
\acro{HSDPA}{High Speed Downlink Packet Access}
\acro{HSPA}{High Speed Packet Access}
\acro{HTTP}{HyperText Transfer Protocol}
\acro{ICMP}{Internet Control Message Protocol} 
\acro{ICI}{Intercell Interference}
\acro{ICIC}{Inter-Cell Interference Coordination}
\acro{ID}{Identification}
\acro{IETF}{Internet Engineering Task Force}
\acro{IPC}{Individual Power Constraint}
\acro{UID}{Unique Identification}
\acro{IID}{Independent and Identically Distributed}
\acro{IIR}{Infinite Impulse Response}
\acro{ILP}{Integer Linear Problem}
\acro{IMT}{International Mobile Telecommunications}
\acro{INV}{Inverted Norm-based Grouping} 
\acro{IoT}{Internet of Things}
\acro{IP}{Internet Protocol}
\acro{IPv6}{Internet Protocol Version 6}
\acro{ISD}{Inter-Site Distance}
\acro{ISI}{Inter Symbol Interference}
\acro{ISM}{Industrial, Scientific and Medical}
\acro{ITU}{International Telecommunication Union}
\acro{JOAS}{Joint Opportunistic Assignment and Scheduling}
\acro{JOS}{Joint Opportunistic Scheduling}
\acro{JP}{Joint Processing}
\acro{JRAPAP}{Joint RB Assignment and Power Allocation Problem}
\acro{JS}{Jump-Stay}
\acro{JSM}{Joint Satisfaction Maximization}
\acro{KKT}{Karush-Kuhn-Tucker}
\acro{KPI}{Key Performance Indicator}
\acro{LAC}{Link Admission Control}
\acro{LA}{Link Adaptation}
\acro{LBS}{Location Based Service}
\acro{LC}{Load Control}
\acro{LOS}{Line of Sight}
\acro{LP}{Linear Programming}
\acro{LTE}{Long Term Evolution}
\acro{LTE-A}{\acs*{LTE}-Advanced}
\acro{LTE-Advanced}{\ac{LTE-A}}
\acro{MeNB}{Master \acs*{eNB}}
\acro{M2M}{Machine-to-Machine}
\acro{MAC}{Medium Access Control}
\acro{MANET}{Mobile Ad hoc Network}
\acro{MEDS}{Method of Exact Doppler Spread}
\acro{MC}{Modular Clock}
\acro{MCP}{Minimal Cost Power}
\acro{MCS}{Modulation and Coding Scheme}
\acro{MDB}{Measured Delay Based}
\acro{MDI}{Minimum \acs*{D2D} Interference}
\acro{MDSM}{Modified Delay-based Satisfaction Maximization}
\acro{MDU}{Max-Delay-Utility}
\acro{METIS}{Mobile and Wireless Communications Enablers for the Twenty-twenty Information Society \acs*{5G}}
\acro{MF}{Matched Filter}
\acro{MG}{Maximum Gain}
\acro{MH}{Multi-Hop}
\acro{MILP}{Mixed Integer Linear Programming}
\acro{MIMO}{Multiple Input Multiple Output}
\acro{MINLP}{Mixed Integer Nonlinear Programming}
\acro{MIP}{Mixed Integer Programming}
\acro{MISO}{Multiple Input Single Output}
\acro{MIT}{Mobility Interruption Time}
\acro{ML}{Machine Learning}
\acro{MLWDF}{Modified Largest Weighted Delay First}
\acro{MME}{Mobility Management Entity}
\acro{MMF}{Max-Min Fairness}
\acro{mmMAGIC}{Millimetre-Wave Based Mobile Radio Access Network for Fifth Generation Integrated Communications}
\acro{MMRP}{Maximizing the Minimal Residual Power}
\acro{MMRP-LB}{Maximizing the Minimal Residual Power with Lower Bound}
\acro{MMSE}{Minimum Mean Square Error}
\acro{mMTC}{Massive Machine-Type Communications}
\acro{mmW}{millimeter Wave}
\acro{MN}{Master Node}
\acro{MOS}{Mean Opinion Score}
\acro{MPF}{Multicarrier Proportional Fair}
\acro{MPRP}{Maximization of the Product of the Residual Powers}
\acro{MRA}{Maximum Rate Allocation}
\acro{MR}{Maximum Rate}
\acro{MRC}{Maximum Ratio Combining}
\acro{MRT}{Maximum Ratio Transmission}
\acro{MRUS}{Maximum Rate with User Satisfaction}
\acro{MS}{Mode Selection}
\acro{MSE}{Mean Squared Error}
\acro{MCG}{Master Cell Group} 
\acro{MSI}{Multi-Stream Interference}
\acro{MTC}{Machine-Type Communication}
\acro{IMS}{\acs*{IP} Multimedia Subsystem}
\acro{MTSI}{Multimedia Telephony Services over \acs*{IMS}}
\acro{MTSM}{Modified Throughput-based Satisfaction Maximization}
\acro{MU-MIMO}{Multi-User Multiple Input Multiple Output}
\acro{MU}{Multi-User}
\acro{Multi-CUT}{Multi-Cell and Multi-User and Multi-Tier}
\acro{NAS}{Non-Access Stratum}
\acro{NB}{Node B}
\acro{NCL}{Neighbor Cell List}
\acro{NGC}{Next Generation Core Network}
\acro{NLP}{Nonlinear Programming}
\acro{NLOS}{Non-Line of Sight}
\acro{NMSE}{Normalized Mean Square Error}
\acro{NN}{Neural Network}
\acro{NOMA}{Non-Orthogonal Multiple Access}
\acro{NORM}{Normalized Projection-based Grouping}
\acro{NP}{Non-Polynomial Time}
\acro{NR}{New Radio}
\acro{NRT}{Non-Real Time}
\acro{NSA}{Non-Stand-Alone}
\acro{NSPS}{National Security and Public Safety Services}
\acro{OBF}{Opportunistic Beamforming}
\acro{OFDMA}{Orthogonal Frequency Division Multiple Access}
\acro{OFDM}{Orthogonal Frequency Division Multiplexing}
\acro{OFPC}{Open Loop with Fractional Path Loss Compensation}
\acro{O2I}{Outdoor-to-Indoor}
\acro{OL}{Open Loop}
\acro{OLPC}{Open-Loop Power Control}
\acro{OL-PC}{Open-Loop Power Control}
\acro{OM}[O\&M]{Operational \& Maintenance}
\acro{OPEX}{Operational Expenditure}
\acro{ORB}{Orthogonal Random Beamforming}
\acro{JO-PF}{Joint Opportunistic Proportional Fair}
\acro{OSI}{Open Systems Interconnection}
\acro{PA}{Power Allocation}
\acro{PAIR}{\acs*{D2D} Pair Gain-based Grouping}
\acro{PAPR}{Peak-to-Average Power Ratio}
\acro{P2P}{Peer-to-Peer}
\acro{PBS}{Pico Base Station}        
\acro{PC}{Power Control}
\acro{PCI}{Physical Cell \acs*{ID}}
\acro{PDCP}{Packet Data Convergence Protocol}
\acro{PDF}{Probability Density Function}
\acro{PDU}{Protocol Data Unit}
\acro{PER}{Packet Error Rate}
\acro{PF}{Proportional Fair}
\acro{P-GW}{Packet Data Network Gateway}
\acro{PHY}{Physical}
\acro{PL}{Pathloss}
\acro{PLR}{Packet Loss Ratio}
\acro{PRABE}{Power and Resource Allocation Based on Quality of Experience}
\acro{PRB}{Physical Resource Block}
\acro{PROJ}{Projection-based Grouping}
\acro{ProSe}{Proximity Services}
\acro{PS}{Packet Scheduling}
\acro{PSO}{Particle Swarm Optimization}
\acro{PTAS}{Polynomial-Time Approximation Scheme}
\acro{PZF}{Projected Zero-Forcing}
\acro{QAM}{Quadrature Amplitude Modulation}
\acro{QHMLWDF}{Queue-HOL-MLWDF}
\acro{QoE}{Quality of Experience}
\acro{QoS}{Quality of Service}
\acro{QPSK}{Quadri-Phase Shift Keying}
\acro{QSM}{Queue-based Satisfaction Maximization}
\acro{QuaDRiGa}{QUAsi Deterministic RadIo channel GenerAtor}
\acro{RACH}{Random Access}
\acro{RAISES}{Reallocation-based Assignment for Improved Spectral Efficiency and Satisfaction}
\acro{RAN}{Radio Access Network}
\acro{RA}{Resource Allocation}
\acro{RAP}{RB Assignment Problem}
\acro{RAT}{Radio Access Technology}[long-plural-form={Radio Access Technologies}]
\acro{RATE}{Rate-based}
\acro{RB}{Resource Block}
\acro{RBG}{Resource Block Group}
\acro{REF}{Reference Grouping}
\acro{RET}{Remote Electrical Tilt}
\acro{RF}{Radio Frequency}
\acro{RL}{Reinforcement Learning}
\acro{RLC}{Radio Link Control}
\acro{RM}{Rate Maximization}
\acro{RMJ-SNR}{Region of the Minimum Joint SNR}
\acro{RMEC}{Rate Maximization under Experience Constraints}
\acro{RNC}{Radio Network Controller}
\acro{RND}{Random Grouping}
\acro{RNN}{Recurrent Neural Network}
\acro{RRA}{Radio Resource Allocation}
\acro{RRM}{Radio Resource Management}
\acro{RSCP}{Received Signal Code Power}
\acro{RSRP}{Reference Signal Received Power}
\acro{RSRQ}{Reference Signal Received Quality}
\acro{RR}{Round Robin}
\acro{RRC}{Radio Resource Control}
\acro{RSSI}{Received Signal Strength Indicator}
\acro{RT}{Real Time}
\acro{RTS}{Request to Send}
\acro{RU}{Resource Unit}
\acro{RUNE}{RUdimentary Network Emulator}
\acro{RV}{Random Variable}
\acro{RZF}{Regularized Zero-Forcing}
\acro{SA}{Subcarrier Assignment}
\acro{SAC}{Session Admission Control}
\acro{sBS}{Serving Base Station}
\acro{SC}{Small Cell}
\acro{SCon}[SC]{Single Connectivity}
\acro{SCG}{Secondary Cell Group}
\acro{SCM}{Spatial Channel Model}
\acro{SCS}{Subcarrier Spacing}
\acro{SC-FDMA}{Single Carrier - Frequency Division Multiple Access}
\acro{SD}{Soft Dropping}
\acro{S-D}{Source-Destination}
\acro{SDPC}{Soft Dropping Power Control}
\acro{SDMA}{Space-Division Multiple Access}
\acro{SMDP}{Semi-Markov	Decision Problem}
\acro{SeNB}{Secondary \acs*{eNB}}
\acro{SER}{Symbol Error Rate}
\acro{SES}{Simple Exponential Smoothing}
\acro{S-GW}{Serving Gateway}
\acro{SINR}{Signal to Interference-plus-Noise Ratio}
\acro{SI}{Satisfaction Indicator}
\acro{SIP}{Session Initiation Protocol}
\acro{SISO}{Single Input Single Output}
\acro{SIMO}{Single Input Multiple Output}
\acro{SIR}{Signal to Interference Ratio}
\acro{SLNR}{Signal-to-Leakage-plus-Noise Ratio}
\acro{SM}{Subcarrier Matching}
\acro{SMA}{Simple Moving Average}
\acro{SN}{Secondary Node}
\acro{SNR}{Signal to Noise Ratio}
\acro{SON}{Self Organizing Networks}
\acro{SORA}{Satisfaction Oriented Resource Allocation}
\acro{SORA-NRT}{Satisfaction-Oriented Resource Allocation for Non-Real Time Services}
\acro{SORA-RT}{Satisfaction-Oriented Resource Allocation for Real Time Services}
\acro{SPF}{Single-Carrier Proportional Fair}
\acro{SRA}{Sequential Removal Algorithm}
\acro{SRB1}{Signaling Radio Bearer~1}
\acro{SRS}{Sounding Reference Signal}
\acro{SSB}{Synchronization Signal Block}
\acro{STTD}{Space Time Transmit Diversity}[long-plural-form={Space Time Transmit Diversities}]
\acro{SU-MIMO}{Single-User Multiple Input Multiple Output}
\acro{SU}{Single-User}
\acro{tBS}{Target Base Station}
\acro{SVD}{Singular Value Decomposition}
\acro{TCP}{Transmission Control Protocol}
\acro{TDD}{Time Division Duplex}
\acro{TDMA}{Time Division Multiple Access}
\acro{TETRA}{Terrestrial Trunked Radio}
\acro{TP}{Transmit Power}
\acro{TPC}{Total Power Constraint}
\acro{TTI}{Transmission Time Interval}
\acro{TTR}{Time-To-Rendezvous}
\acro{TTT}{Time-To-Trigger}
\acro{TSM}{Throughput-based Satisfaction Maximization}
\acro{TU}{Typical Urban}
\acro{TV}{Television}
\acro{TVWS}{\acs*{TV} White Space}
\acro{UDP}{User Datagram Protocol}
\acro{UABSC}{User-Assisted Bearer Split Control}
\acro{UE}{User Equipment}
\acro{UBA}{\ac{UE}-\ac{BS} association}
\acro{UEPS}{Urgency and Efficiency-based Packet Scheduling}
\acro{UFC}{Federal University of Cear\'{a}}
\acro{ULA}{Uniform Linear Array}
\acro{UL}{Uplink}
\acro{UMTS}{Universal Mobile Telecommunications System}
\acro{URA}{Uniform Rectangular Array}
\acro{URI}{Uniform Resource Identifier}
\acro{URLLC}{Ultra-Reliable Low-Latency Communications}
\acro{URM}{Unconstrained Rate Maximization}
\acro{V2V}{Vehicle-to-Vehicle}
\acro{VBR}{Variable Bit Rate}
\acro{VET}{Variable Electrical Tilt}
\acro{VR}{Virtual Resource}
\acro{VoIP}{Voice over \acs*{IP}}
\acro{WCDMA}{Wideband Code Division Multiple Access}
\acro{WF}{Water-filling}
\acro{Wi-Fi}{Wireless Fidelity}
\acro{WiMAX}{Worldwide Interoperability for Microwave Access}
\acro{WINNER}{Wireless World Initiative New Radio}
\acro{WLAN}{Wireless Local Area Network}
\acro{WMPF}{Weighted Multicarrier Proportional Fair}
\acro{WP}{Work Package}
\acro{WPF}{Weighted Proportional Fair}
\acro{WSN}{Wireless Sensor Network}
\acro{WWW}{World Wide Web}
\acro{WFQ}{Weighted Fair Queuing}
\acro{XIXO}{(Single or Multiple) Input (Single or Multiple) Output}
\acro{ZF}{Zero-Forcing}
\acro{ZMCSCG}{Zero Mean Circularly Symmetric Complex Gaussian}

\usepackage{siunitx}
\usepackage{datetime}
\usepackage{url}

\usepackage[caption=false,font=footnotesize]{subfig}

\usepackage{algorithm}
\usepackage{algorithmicx}
\usepackage{algpseudocode}
\usepackage{epstopdf}

\usepackage[hidelinks=true]{hyperref}

\usepackage[draft]{changes}

\definechangesauthor[name={Victor F. Monteiro}]{vfm}
\setaddedmarkup{{\color{blue}#1}}
\setdeletedmarkup{{\color{red}\sout{#1}}}

\usepackage[referable,flushleft]{threeparttablex}
\AtBeginEnvironment{tablenotes}{\footnotesize} %

\renewcommand*{\bibfont}{\footnotesize}

\DeclareMathAlphabet{\mathppl}{T1}{ppl}{m}{it}
\DeclareMathAlphabet{\mathphv}{T1}{phv}{m}{it}
\DeclareMathAlphabet{\mathpzc}{T1}{pzc}{m}{it}

\newcommand{\Mt}[1]{\mathbf{#1}}

\newcommand{\Set}[1]{\mathcal{\uppercase{#1}}}

\newcommand{\Vt}[1]{\mathbf{\lowercase{#1}}}

\newcommand{\mtX}{\Mt{X}}

\newcommand{\vtX}{\Vt{X}}

\newcommand{\vtRho}{\Vt{\boldsymbol{\rho}}}

\newcommand{\stD}{\Set{D}}

\newcommand{\SecRef}[2][]{Section#1~\ref{#2}}
\newcommand{\FigRef}[2][]{Fig.#1~\ref{#2}}

\newcommand{\TabRef}[2][]{Table#1~\ref{#2}}

\newcommand{\AlgRef}[2][]{Alg.#1~\ref{#2}}%

\usepackage{tikz}
\usetikzlibrary{arrows}
\usetikzlibrary{positioning}
\usetikzlibrary{shapes.misc}
\usetikzlibrary{shapes.geometric}
\usetikzlibrary{shapes.symbols}
\usetikzlibrary{external}
\tikzset{external/system call={pdflatex \tikzexternalcheckshellescape -halt-on-error -interaction=batchmode -jobname "\image" "\texsource"}}

\usepackage{pgfplots}
\pgfplotsset{%
	height = 0.17\textheight, %
	width=\columnwidth,%
	compat=1.14,
	compat/show suggested version=false,
	filter discard warning=false,
	tick label style={font=\footnotesize},
	label style={font=\footnotesize},
	every axis label={font=\footnotesize},
	grid=major,
	grid style={dashed,gray!30},
	cycle list shift=0,
	enlargelimits=false,
	legend style={%
		font=\footnotesize,
		legend cell align=left,
		nodes={inner xsep=2pt,inner ysep=1pt,text depth=0.15em},
	},
}

\AtBeginDocument{%
	\abovedisplayskip 1.5ex plus 4pt minus 2pt
	\belowdisplayskip \abovedisplayskip
	\abovedisplayshortskip 0pt plus 4pt
	\belowdisplayshortskip 1.5ex plus 4pt minus 2pt
}

\usepackage{float}
\begin{document}

\definecolor{colorSysThroughput}{RGB}{255,164,5}
\definecolor{colorUE1}{RGB}{153,63,0}
\definecolor{colorUE2}{RGB}{120,120,120}

\definecolor{colorOptimal}{RGB}{255,164,5}
\definecolor{colorRAISES}{RGB}{153,63,0}
\definecolor{colorRMEC}{RGB}{120,120,120}
\definecolor{colorProposal1k}{RGB}{0,180,100}
\definecolor{colorProposal3k}{RGB}{0,51,128}

\tikzset{plot common style/.append style={line width=1pt, mark options={scale=1pt, solid}}}
\tikzset{SysThroughput style/.append style={plot common style, solid, mark=square*, color=colorSysThroughput}}
\tikzset{UE1 style/.append style={plot common style, solid, mark=square*, color=colorUE1}}
\tikzset{UE2 style/.append style={plot common style, solid, mark=square*, color=colorUE2}}

\tikzset{Optimal style/.append style={plot common style, solid, mark=square*, color=colorOptimal}}
\tikzset{RAISES style/.append style={plot common style, solid, mark=diamond*, color=colorRAISES}}
\tikzset{RMEC style/.append style={plot common style, solid, mark=asterisk, color=colorRMEC}}
\tikzset{Proposal1k style/.append style={plot common style, solid, mark=o, color=colorProposal1k}}
\tikzset{Proposal3k style/.append style={plot common style, solid, mark=triangle*, color=colorProposal3k}}

\def\bitsVSepisodes{figs/Results_dat/bitsVSepisodes.dat}
\def\outageTable{figs/Results_dat/outage.txt}
\def\sysThroughputTable{figs/Results_dat/sysThroughput.txt}

\title{Deep Reinforcement Learning for QoS-Constrained Resource Allocation in Multiservice Networks}

\author{Juno V. Saraiva, Iran M. Braga Jr.,
        Victor F. Monteiro, F. Rafael M. Lima, \\Tarcisio F. Maciel, Walter C. Freitas Jr. and F. Rodrigo P. Cavalcanti%
\thanks{This work was supported by Ericsson Research, Sweden, and Ericsson Innovation Center, Brazil, under UFC.46 and UFC.47 Technical Cooperation Contracts Ericsson/UFC. Iran M. Braga Jr. would like to acknowledge CAPES for its financial support under	the grant 88887.474363/2020-00.}%
\thanks{The authors are with the Wireless Telecommunications Research Group (GTEL), Federal University of Cear\'{a} (UFC), Fortaleza, Cear\'{a}, Brazil.}%
\thanks{} }

\maketitle
\begin{abstract}
In this article, we study a Radio Resource Allocation (RRA) that was formulated as a non-convex optimization problem whose main aim is to maximize the spectral efficiency subject to satisfaction guarantees in multiservice wireless systems. This problem has already been previously investigated in the literature and efficient heuristics have been proposed. However, in order to assess the performance of Machine Learning (ML) algorithms when solving optimization problems in the context of RRA, we revisit that problem and propose a solution based on a Reinforcement Learning (RL) framework. Specifically, a distributed optimization method based on multi-agent deep RL is developed, where each agent makes its decisions to find a policy by interacting with the local environment, until reaching convergence. Thus, this article focuses on an application of RL and our main proposal consists in a new deep RL based approach to jointly deal with RRA, satisfaction guarantees and Quality of Service (QoS) constraints in multiservice celular networks. Lastly, through computational simulations we compare the state-of-art solutions of the literature with our proposal and we show a near optimal performance of the latter in terms of throughput and outage rate.
\end{abstract}

\begin{IEEEkeywords}
Radio resource allocation, QoS, satisfaction guarantees,
machine learning, reinforcement learning, deep neural network, deep $Q$-learning.%
\end{IEEEkeywords}

\IEEEpeerreviewmaketitle
\acresetall

\section{Introduction} \label{SEC:Intro}
The overall performance of wireless telecommunications systems directly depends on how efficiently the available resources are managed, e.g., subcarriers, time slots, transmit power, antennas, among others. %
Consequently, optimal \ac{RRA} is one of the fundamental challenges and a key requirement for the design of efficient mobile networks. \ac{RRA} problems in general are formulated as optimization problems and different objective functions and constraints have been considered in the literature. %
Two examples of classical objective functions are: maximize the system throughput, as considered in~\cite{Lima2012} and~\cite{Sousa2016} and guarantee system fairness, as done in~\cite{Monteiro2019}. However, with the advent of the \ac{5G} of mobile wireless telecommunications, which shall integrate new technology components, \ac{RRA} problems can be come even harder to tackle, with larger optimization domains and a series of practical considerations. Moreover, these problems need also to deal with the advanced \ac{RRA} functionalities, the growing variety of scenarios and sophisticated types of services and/or \ac{QoS} constraints of the users~\cite{Calabrese2018}.

Although it is possible to apply optimal methods to some of these problems, such as exhaustive search and \ac{BB} algorithm, their high computational complexities are prohibitive and, therefore, these methods are not appealing for large-scale mobile networks. Contrarily, techniques such as Lagrangian relaxations, iterative distributed optimization and heuristic algorithms normally have reduced computational costs,
but they fail in achieving the maximum performance and are usually tailored to specific network configurations. Furthermore, issues related to convergence and optimality gaps of these solutions can be unknown as well~\cite{Ahmed2019}. As a result, the solution of many problems in the context of \ac{RRA} can be quite inadequate using conventional optimization methods. There already exists a very rich literature on this topic, as it can be seen in~\cite{Castaneda2017}, where an extensive survey on these techniques is presented.

\ac{ML} techniques are leading the advent of the fourth industrial revolution due to its capabilities of solving complex problems. In fact, this has aroused interest of many researchers in the mobile communications field. More specifically, in learning-based resource allocation, a branch of \ac{ML} called deep learning has gained notoriety and shown its potential in this type of context. Moreover, in order to further improve the performance of this technique with regard to learning from high-dimensional raw input data and make intelligent decisions, in~\cite{Mnih2015}, it is proposed a sophisticated approach where deep learning and \ac{RL} are combined, resulting in a promising and powerful technique known as deep \ac{RL}. In summary, in deep \ac{RL}, a \ac{DNN} along with other techniques, e.g., replay memory, are used in order to carry out a stable and efficient training.

Applying deep \ac{RL} to cellular mobile networks can lead to the following main advantages: (1) a \ac{DNN}  with a moderate size can quickly perform predictions as only a small number of simple operations are needed to obtain an output. This is interesting and also helps the deep \ac{RL} agent to get to know his environment faster; (2) the fact of the deep RL agent learn directly from the raw collected network data with high dimension in large environments is not a problem due to the powerful representation capabilities of \acp{DNN}; (3) by exploiting distributed and/or parallel computing employing multiple machines and multiple cores, the response time of deep \ac{RL}-based schemes can be greatly reduced and its performance increased; (4) deep \ac{RL}-based schemes can also improve over time since the  deep \ac{RL} agent aims at optimizing a long-term  performance, considering the impact of actions on future rewards. This makes deep \ac{RL} efficient in dealing with imprecise input data such as the \ac{CSI} and makes it
 capable of learning how to behave in an unknown environment~\cite{Mao2018}; (5) deep \ac{RL} is a model-free approach, i.e., it does not rely on a specific system  model and, therefore, it can be easily extended to different contexts~\cite{Yaohua2019}.

Motivated by the benefits of deep \ac{RL}, in this paper, we revisit the problem of maximizing system throughput  subject to minimum satisfaction constraints
per service, as in \cite{Lima2012} and~\cite{Sousa2016}, and we propose a new near-optimal \ac{RRA} solution based on multiple agent deep \ac{RL}. %

The remainder of this paper is organized as follows. %
In \SecRef{SEC:StateOfArt} previous related works are reviewed, and the main contributions of our work are highlighted. %
In \SecRef{SEC:SysMod} and \SecRef{SEC:ProbForm}, we present the system modeling and formulate the optimization problem to be solved, respectively. %
\SecRef{SEC:ProposedSol} presents a deep \ac{RL}-based method to solve the problem formulated in the previous section. %
\SecRef{SEC:PerfEval} and \SecRef{SEC:Conclusion} show simulation results and the main conclusions of this study, respectively. %

\section{State-of-the-Art and Main Contributions} \label{SEC:StateOfArt}
\ac{RL} techniques have recently been used in a variety of wireless resource management problems such as, channel and power allocation, throughput maximization and spectrum sharing. In~\cite{Amiri2018}, for example, a self-organizing method to allocate power to \ac{mmW} \acp{BS} is proposed based on $Q$-learning technique. $Q$-learning is the most popular \ac{RL} algorithm, where an agent interacts over time with its environment based on trial-and-error in order to learn a policy to achieve a given goal~\cite{Sutton2018}. %
Thus, based on this technique, in \cite{Amiri2018}, each \ac{BS} acts as an independent agent taking actions such as choosing a transmit power. In other words, each \ac{BS} sees the others as part of the environment and do not communicate with each other. In this case, the environment, for a given \ac{BS}, is seen as a source of interfering signals. %
The problem with this solution is that, in a non stationary environment as mobile wireless networks, it takes a long time to converge, due to the lack of cooperation between the \acp{BS}.

$Q$-learning based resource allocation models are also proposed in \cite{Romero2016} and \cite{Saraiva2019}. Specifically, in~\cite{Romero2016}, $Q$-learning is used to select which network node a \ac{UE} should connect to in order to minimize the total transmit power. %
It is considered that a \ac{UE} can either directly connect to a \ac{BS} or to another \ac{UE}, which relays the traffic from a \ac{BS}. %
According to that solution, each \ac{UE} autonomously selects the node to which it will be connected and keeps a record of its experience when using that node. %
The records, called rewards, reflect the degree of fulfillment of the optimization target, e.g., total transmit power used by \acp{BS} and also other constraints such as required bit rate. %
In~\cite{Saraiva2019}, we have proposed a $Q$-learning based solution to the same problem that we address in the present paper, i.e., schedule frequency resources to \acp{UE} in order to maximize the system throughput subject to users' \ac{QoS} requirements, in terms of \ac{UE} throughput. However, in general, $Q$-learning based solutions, and especially the one proposed in~\cite{Saraiva2019}, present a scalability problem, since the agent's experiences are stored in a look-up table, called $Q$-table. %
This becomes an issue when the set of possible experiences increases with the dimensions of the environment, which is the case in~\cite{Saraiva2019}, where the size of the $Q$-table is proportional to the number of frequency resources and the number of \acp{UE}. %

As presented in the previous section, when a \ac{DNN} is used by an agent, it is referred to as deep \ac{RL} and this technique has shown to be efficient for \ac{RRA} in large cellular networks. In~\cite{Nasir2019}, for example, a model-free deep \ac{RL} method is applied to perform dynamic transmit power allocation. %
The solution presented in that work achieves near-optimal performance and is suitable for practical scenarios where the system model is inaccurate, thus overcoming some issues of classical and heuristic solutions. A decentralized band and power allocation problem for a  \ac{V2V} communication system is solved based on deep \ac{RL} in \cite{Hao2019}. In details, the goal is to minimize interference under latency constraints, where each \ac{V2V} link operates as an agent making its own allocation decisions. Moreover, many other papers also successfully address deep \ac{RL} in several wireless telecommunications research areas, such as  resource scheduling~\cite{Zhao2019} and mobile edge computing/caching~\cite{He2017, He2017Deep}.%

The above mentioned works consider either a completely centralized solution \cite{He2017, He2017Deep} or a decentralized solution \cite{Nasir2019, Hao2019, Zhao2019}. %
On one hand, in a centralized solution, a deep \ac{RL} agent is localized in a central node and it is responsible for the entire processing. Unfortunately, this processing consumes a lot of computational resources since the employed \ac{DNN} size is proportional to the wireless network dimension. %
On the other hand, in a decentralized solution, multiple agents are considered each one running in a different node and taking actions independently. %
In this last option, the agents can either work independently or in cooperation with the burden of a longer convergence time or a higher signalling overhead and data updating procedures, respectively. %
In the present work, we propose another option, where we assume a centralized node, but with multiple agents running in parallel, each one related to a frequency resource. In addition, this structure is executed independently in a distributed way in different cores or machines in order to improve the performance of the system. %

In summary, our main contributions are the following:
\begin{enumerate}
	\item We revisit an important \ac{RRA} problem originally studied with the help of optimization tools and heuristics and propose a new methodology that leverages decentralized deep \ac{RL}. This methodology could be applied in other RRA problems to obtain alternative solutions. To the best of our knowledge, this strategy has not yet been applied to frequency resource scheduling problems.
	\item By means of extensive computer simulations, we show that our proposed deep \ac{RL}-based solution outperfoms the state-of-art solutions found in the literature. Indeed, this is interesting because it shows that our deep \ac{RL} approach is capable of dealing with a challenging scenario of mobile networks that includes satisfaction of users' QoS in different types of service plans, something that few papers consider in the literature. %
\end{enumerate}

\section{System Modeling} \label{SEC:SysMod}
We consider a \ac{SISO} downlink cellular system composed of a number of sectored cells so that in a given sector there are $J$ \acp{UE} grouped in set $\mathcal{J}=\{1,\dots,J\}$ and connected to an \ac{eNB}. %

We assume that the intra-cell interference, i.e., interference between terminals of the same cell, is controlled by employing the combination
of \ac{OFDMA} and \ac{TDMA} with the assignment of orthogonal resources. Thus, we define a \ac{RB} as the basic scheduling unit composed of a group of subcarriers in the frequency domain and a number of consecutive \ac{OFDM} symbols in the time domain, whose total duration represents a \ac{TTI}. In addition, $\mathcal{N} = \{1, \dots, N\}$ is the set of \acp{RB} available. Regarding inter-cell interference, we assume that it is added to the thermal noise in the \ac{SNR} expression, defined later.

In this work, we also assume a multiservice scenario with $L$ service plans contained in the set $\mathcal{L}=\{1, \dots, L\}$ and supported by the system operator. %
In each \ac{TTI}, the $J$ \acp{UE} compete for the available \acp{RB} in order to meet their throughput requirements, which are defined by their service plans. %
Each service plan $l \in \mathcal{L}$ requires a minimum number of \acp{UE} that should be satisfied. %
The set of all \acp{UE} from service $l \in \mathcal{L}$ is $\mathcal{J}_{l}$ with $|\mathcal{J}_{l}|=J_{l}$, where $|\cdot|$ denotes the cardinality of a set and $\mathcal{J}_{l}$ is the set of \acp{UE} from service $l \in \mathcal{L}$. %
Besides, each \ac{UE} subscribes to only a single service plan, i.e., $\mathcal{J}_{l_{1}}\cap \mathcal{J}_{l_{2}}=\emptyset$, $\forall l_{1},l_{2} \in \mathcal{L}$ and $l_{1}\neq l_{2}$. %

The \ac{SNR} $\gamma_{j,n}$ of \ac{UE} $j\in \mathcal{J}$ in \ac{RB} $n \in \mathcal{N}$ is given by

\begin{equation}
\label{EQ:POT_MIN}
\gamma_{j,n} = \frac{p_{n}\cdot \alpha_{j} \cdot |h_{j,n}|^{2}}{\sigma^{2}},
\end{equation} %
where $p_{n}$ is the transmit power allocated to the \ac{UE} $j$ on \ac{RB} $n$; %
$\alpha_{j}$ models the joint effect of the path loss and shadowing of the link between the \ac{eNB} and \ac{UE} $j$; %
$|h_{j,n}|$ represents the magnitude of the complex channel frequency response of \ac{RB} $n$ when assigned to \ac{UE} $j$; %
and, finally, $\sigma^{2}$ is the noise power at the receiver in the bandwidth of a given \ac{RB}. %

Similar to~\cite{Lima2012},~\cite{Sousa2016} and~\cite{Saraiva2019}, power allocation is not optimized herein and we employ \ac{EPA} among \acp{RB}, which is the most basic and common power allocation scheme. %
Hence, the power $p_{n}$ allocated to each RB $n$ is fixed and equal to $P/N$, where $P$ is the available power at the \ac{eNB}. %

We assume $f(\cdot)$ as the link adaptation function responsible for mapping the achieved \ac{SNR} to the transmit rate. %
It is a discrete and monotonic increasing function that models the \ac{MCS} levels so that the transmission parameters at the physical layer are adapted according to the current channel state. %
Thus, we consider that the transmit rate when the \ac{RB} $n$ is assigned to \ac{UE} $j$ is $r_{j,n}$ such that $r_{j,n}=f(\gamma_{j,n})$. %

\section{Problem Formulation and Optimal Solution} \label{SEC:ProbForm}
As presented in Section \ref{SEC:StateOfArt}, the problem investigated herein aims to maximize the system throughput constrained by a per-service minimum number of satisfied \acp{UE} in a given \ac{TTI}. %
For that problem, we define $x_{j,n}$ as the binary decision variable that assumes the value $1$ when \ac{RB} $n$ is assigned to \ac{UE} $j$ and $0$, otherwise. %
Furthermore, let $R_{j}$ be the total throughput allocated to a \ac{UE} $j$, i.e., $R_{j}=\sum_{n \in \mathcal{N}} r_{j,n}x_{j,n}$. %
Therefore, the resource assignment problem can be formulated as:

\begin{subequations}
	\label{EQ:OBJ2}
	\begin{align}
	&\max_{\mtX} \sum_{j\in \mathcal{J}} R_{j} , \label{EQ:Objective_function}\\
	\text{s.t. }
	\label{CONS:1}
	&\sum_{j \in \mathcal{J}}x_{j,n} = 1, \,\, \forall n \in \mathcal{N},\\
	\label{CONS:NONLINEAR}
	&\sum_{j\in \mathcal{J}_{l}}u\left( R_{j}, \xi_{j}\right) \geq \eta_{l}, \,\, \forall l \in \mathcal{L},\\
	\label{CONS:3}
	& x_{j,n} \in \{0,1\}, \,\, \forall j \in \mathcal{J}\text{ and } \forall n \in \mathcal{N},
	\end{align}
\end{subequations} %
where $\mtX$ is the matrix of optimization variables composed of $x_{j,n}$ and $u(a,b)$ in \eqref{CONS:NONLINEAR} denotes the Heaviside step function, which assumes the value $1$ if $a \geq b$ and $0$, otherwise. %
With this, $\eta_{l}$ is the minimum number of \acp{UE} from service $l$ that should be satisfied and $\xi_{j}$  represents the required throughput for a \ac{UE} to be considered satisfied, i.e., $\xi_{j}$ consists of a \ac{QoS} requirement for each \ac{UE} $j$ in terms of throughput. %
Regarding constraints \eqref{CONS:1} and \eqref{CONS:3}, they guarantee that each \ac{RB} is assigned to a single \ac{UE}. %
Notice that \eqref{EQ:OBJ2} is a combinatorial optimization problem with a non-convex constraint \eqref{CONS:NONLINEAR}. In order to simplify the optimal solution analyses, we linearize \eqref{CONS:NONLINEAR} by introducing some new variables. %
Let $\rho_{j}$ be a binary selection variable that assumes the value $1$ if \ac{UE} $j$ is selected to be satisfied and $0$, otherwise. %
Thus, \eqref{CONS:NONLINEAR} can be replaced by $\eqref{CONS:3_new}$ and $\eqref{CONS:4_new}$, where $\rho_{j}=1$ in \eqref{CONS:3_new} implies that \ac{UE} $j$ is satisfied and $\eqref{CONS:4_new}$ means that for all service $l$ there are at least $\eta_{l}$ satisfied \acp{UE}. %
Hence, problem \eqref{EQ:OBJ2} can be equivalently reformulated as

\begin{subequations}
	\label{EQ:OBJ3}
	\begin{align}
	&\max_{\vtX,\vtRho} \sum_{j\in \mathcal{J}} R_{j}  ,\\
	\text{s.t. }
	&\sum_{j \in \mathcal{J}}x_{j,n} = 1, \,\, \forall n \in \mathcal{N},\\
	\label{CONS:3_new}
	& R_{j} \geq \xi_{j}\cdot \rho_{j}, \,\, \forall j \in \mathcal{J},\\
	\label{CONS:4_new}
	&\sum_{j \in \mathcal{J}_{l}}\rho_{j} \geq \eta_{l}, \,\, \forall l \in \mathcal{L},\\
	& x_{j,n},\,\, \rho_{j} \in \{0,1\}, \,\, \forall j \in \mathcal{J}\text{ and } \forall n \in \mathcal{N}.
	\end{align}
\end{subequations} %

Thus, we have transformed \eqref{EQ:OBJ2} into an \ac{ILP}, which can be solved by standard methods such as the \ac{BB} algorithm \cite{Lima2012,Sousa2016}.

\section{Proposed solution} \label{SEC:ProposedSol}
In this section, in order to better understand our proposed solution, a brief review of \ac{RL}, including the techniques $Q$-learning and deep \ac{RL}, is first described.
\subsection{An Overview of Reinforcement Learning} \label{SUBSEC:Overview_reinforcement-learning}
\subsubsection{$Q$-Learning technique} \label{SUBSEC:Q-learning}

One of the most common RL techniques is the $Q$-learning model which consists of an \textit{agent}, a set of \textit{states} $\mathcal{S}$ and a set of \textit{actions} per state $\mathcal{A}(s)$, $s\in \mathcal{S}$. %
By performing actions and, consequently, transitions from state to state, the agent aims to learn an optimal policy or an optimal path to a given goal. %
Each of these states can be defined as a tuple of values that characterizes the environment for the agent, while each action represents the change that the agent applies to this environment. %
Thereby, the idea is that the agent perceives the environment state and selects an action according to a particular strategy or decision policy~\cite{Sutton2018}. %
This strategy or policy can be implemented using a variety of techniques such as the $\epsilon$-greedy decision policy. %
Particularly, the $\epsilon$-greedy policy is simple, but very efficient: the agent \textit{explores} or \textit{exploits} its environment taking random (non-greedy) or greedy actions according to a given probability distribution, respectively. %
Normally, for this decision policy, a random action can be chosen with probability $\epsilon \in (0,1)$, while a greedy action is taken with a probability $1-\epsilon$. %

On one hand, when greedy actions are taken, the objective is to exploit the acquired knowledge to improve performance. %
Generally, for a (partially-)unknown environment, these actions lead to locally optimal solutions. On the other hand, when the agent decides to take random actions, it explores its environment for the sake of acquiring experience and knowledge about the environment and, therefore, there is no concern with the immediate effects of these actions. %
However, random actions allow the agent to neglect the locally optimal policies, and to achieve the globally optimal one, instead. %
Consequently, one of the challenges that arise in \ac{RL} techniques is the trade-off between \textit{exploration} and \textit{exploitation} and it should be carefully balanced so that the benefits of both can be properly harvested~\cite{Sutton2018}.

Once taken an action $a\in \mathcal{A}(s)$, the system state changes from $s$ to $s'$ and this change generates a signal or indicator that evaluates the effect of the taken action. %
This feedback or message from the environment is called \textit{reward}, $\phi$, which is a numerical score and it is used to estimate the expected value of taking an action $a$ in a particular state $s$, also known as $Q$-value of a state/action $(s, a)$. %
In detail, the $Q$-value is calculated by a $Q$-function such that $Q:\mathcal{S}\times \mathcal{A}\rightarrow \mathbb{R}$ and, for a given state/action pair $(s,a)$, it can be estimated according to Bellman's equation~\cite{Sutton2018}:
\begin{equation}
\label{EQ:BELLMAN}
\hat{Q}(s,a) = (1-\alpha)\hat{Q}(s,a)+\alpha(\phi+\gamma \max_{a'}\hat{Q}(s',a')),
\end{equation}
where $0<\alpha \leq 1$, $0\leq \gamma < 1$ are constants called learning rate and discount factor, respectively, and $\max_{a'}\hat{Q}(s',a')$ is the best estimated $Q$-value given the next state $s'$ and all possible actions at $s'$. %
Basically, $\alpha$ determines how quickly the learning process occurs, while $\gamma$ controls the value placed on future $Q$-values. %
Then, over several iterations the state/action pairs are defined and their respective $Q$-values are estimated and updated by Bellman's equation. A set of these iterations from an \textit{initial state} $s_{\text{o}}$ to a \textit{final state} $s_{\text{f}}$ is called an \textit{episode}. %

Thereby, each state/action pair and its respective performed action allow the agent to interact with the environment and this interaction after several episodes produces precious information about the consequences of actions, mainly about what to do or not in order to achieve goals. %
This information are precisely represented in the $Q$-values and the set of all of them is stored in the $Q$-table, which is where all the experience or knowledge acquired by the agent is stored. Therefore, the basic idea in $Q$-learning is that the agent finds and learns an optimal policy for the desired problem by benefiting from the experience gathered in the $Q$-table.

\subsubsection{Deep $Q$-Learning technique} \label{SUBSEC: Deep Q-learning}
Although the $Q$-learning technique is very simple, it is a quite powerful algorithm to create an interesting set of experience or a kind of cheat-sheet for the agent. Indeed, this is fundamental and helps the agent to figure out exactly which action to perform until it converges to an optimal policy. Nevertheless, as highlighted in~\cite{Saraiva2019}, $Q$-learning has two serious problems: (1) the amount of memory required to save and update the $Q$-table can increase exponentially as the number of states and actions increases, (2) many states are rarely visited and, consequently, the amount of time required to explore all these possibilities (state/action pairs) in order to create a good estimate for $Q$-table would be unrealistic or impractical in a real setting \cite{Nasir2019}.

As a result, producing and updating a $Q$-table can become ineffective in large-sized environments, i.e., with a large number of states and actions. Notwithstanding, these limitations can be solved with the emerging deep \ac{RL}, e.g., deep $Q$-learning, which is considered as a promising technique to solve the complex control issues, especially for the high-dimension solutions \cite{Zhao2019}. Basically, this technique can be cast as an extension of classical $Q$-learning algorithm that uses \ac{DNN} to approximate the $Q$-function in lieu of a lookup table.

In specific, in the deep $Q$-learning algorithm, a \ac{DNN} called \ac{DQN} is defined as a parameterized value function $Q_{\mat{\theta}}:\mathcal{S}\times \mathcal{A}\rightarrow\mathbb{R}$ that is used to estimate the $Q$-function, where the state is given as its input, the $Q$-value of all possible actions is generated as its output and, finally, $\mat{\theta}$ represents its parameters that define the $Q$-values. Therefore, the key idea of deep $Q$-learning technique is that the function $Q_{\mat{\theta}}$ is completely determined by $\mat{\theta}$. Consequently, the task of finding the best $Q$-function is essentially limited to the search for these best parameters of finite dimensions \cite{Nasir2019}.

Although the algorithmic and statistical properties as well as the performance of the classical $Q$-learning algorithm are well known and studied, the same is not true for deep $Q$-learning, which still remains less well-understood in theory. Thereby, the idea of simply approximating the $Q$-function by a \ac{DNN} can often lead to learning instability. Fortunately, these instabilities can be greatly reduced by the following two aspects \cite{Franccois2018,Yang2019}. Firstly, similar to classical $Q$-learning, the agent interacts with its environment following an $\epsilon$-greedy policy over $\mathcal{S}\times\mathcal{A}$. However, the experiences, composed each one of them of the current state ($s$), action ($a$), reward ($\phi$), and
the next state ($s'$), obtained by agent are gathered in a memory $\mathcal{D}$ with limited capacity $D$. In addition, the DQN training is performed on a mini-batch $\mathcal{B}$ of $B$ tuples or experiences selected randomly from $\mathcal{D}$. Such method is known as \textit{memory replay}. This strategy can  reduce the correlation among the training examples, which ensures that the optimal policy cannot be driven to a local minimal \cite{Yang2019}.

Moreover, when only a single \ac{DQN} is employed, the same values are used to select and evaluate an action and, thereby, the $Q$-function may be over-optimistically estimated. Therefore, the second important aspect consists of using two \acp{DQN} with the same architecture: the target \ac{DQN}, $Q_{\mat{\theta}_{\text{target}}}$, with parameters $\mat{\theta}_{\text{target}}$ and the training \ac{DQN}, $Q_{\mat{\theta}_{\text{train}}}$, with parameters $\mat{\theta}_{\text{train}}$. The training \ac{DQN} is responsible for learning the values of $\mat{\theta}_{\text{train}}$, while the target \ac{DQN} is used to take actions. The value of $\mat{\theta}_{\text{target}}$ are updated at every $\tau$ iterations and set to be equal to $\mat{\theta}_{\text{train}}$. Put another way, the weights $\mat{\theta}_{\text{target}}$ are fixed for a number of iterations while the weights $\mat{\theta}_{\text{train}}$ are constantly updated. This strategy is commonly called Double DQN (DDQN) and it has shown improvements in learning process compared to single \ac{DQN} strategy \cite{Hasselt2016}. Therefore, for each episode, the least squares loss of the training \ac{DQN} for a random mini-batch $\mathcal{B} \subset \stD$ with $B$ samples is

\begin{equation}
\label{L}
\Omega(\mat{\theta}_{\text{train}})=\sum_{(s,a,\phi,s')\in\mathcal{B}} (y - Q_{\mat{\theta}_{\text{train}}}(s,a))^{2},
\end{equation}
where the target is
\begin{equation}
\label{L!}
y = \phi + \gamma \max_{\forall a'} Q_{\mat{\theta}_{\text{target}}}(s',a').
\end{equation}

Finally, the loss function \eqref{L} is minimized by a stochastic gradient algorithm in order
 to train the mini-batch $\mathcal{B}$. Then, the train DQN updates its parameters with the new
 parameters provided by training. According to \cite{Nasir2019}, the convergence to a set of good parameters
occurs quickly.

\subsection{Proposed Multi-Agent Deep $Q$-learning Solution} \label{SUBSEC:ProposedSol}
In  this  section,  we  present  a  multi-agent  DQN-based dynamic resource allocation framework to solve problem \eqref{EQ:OBJ2}. Firstly, we define the concept of agent, state, action, and reward for this approach.

\begin{itemize}
	\item \textbf{Agents}: we propose a multi-agent deep reinforcement learning scheme with each RB as an agent. As a result, there are $N$ agents in this approach.
	\item \textbf{Action of agent} $n$: consist of choosing a UE $j$, i.e., an action $a^{(n)}=j$ means that RB $n$ is assigned to UE $j$. A tuple or vector $\mathbf{a}$, composed of elements $a^{(n)}$, therefore, means a given assignment pattern or association among \acp{UE} and \acp{RB}.
	\item \textbf{State of agent $n$}:  we describe the state of agent $n$, $s^{(n)}$, as a composition of two important aspects for an agent. In the first part, we consider a piece of information common to all agents. This information consists in an $N$-tuple $\mathbf{a}$ which represents a possible assignment for the system. On the other hand, in the second part, we have specific information related to agent $n$. Thus, the second part of the state of agent $n$ is composed by a $J$-tuple, $\mathbf{u}$, where $u^{(n)}_{j}=\gamma_{j,n}, \,\, \forall j$, i.e., each agent or RB $n$ knows the SNR value for all users of the system.
	\item \textbf{Reward}: obviously, the reward function should be designed to maximize the objective  \eqref{EQ:Objective_function} of problem \eqref{EQ:OBJ2}. Thus, to do that we use \AlgRef{ALG:MET2}. The main idea of this algorithm is to define a reward value, $\phi$, capable of reporting what is possible to achieve in terms of satisfaction and system throughput for a given assignment. %
	This value tries to measure how close one is from meeting the requirements of problem \eqref{EQ:OBJ2}, without disregarding its objective function. %
	Note that if all constraints of problem \eqref{EQ:OBJ2} are met, meaning that the chosen assignment is a feasible solution of problem \eqref{EQ:OBJ2}, then $\phi$ is equal to $\sum_{j \in \mathcal{J}}R_{j}$, which is the objective function \eqref{EQ:Objective_function}. %
	Otherwise, $\phi$ is equal to $\vartheta / \sum_{j \in \mathcal{J}}R_{j}$. %
	The variable $\vartheta$ is responsible for quantifying how close the chosen assignment is to a feasible solution of problem \eqref{EQ:OBJ2}. %
	Notice that if any constraint is not met, then $\vartheta$ is negative, and this represents a punishment or a negative reward.
\end{itemize}

\begin{algorithm}[t]
	{\scriptsize
		\caption{\small Set reward value}
		\footnotesize
		\begin{algorithmic}[1]
			\Require $R_{j} \text{ and } \xi_{j}, \,\, \forall j \in \mathcal{J}$;
			\State $\phi \leftarrow \sum_{j \in \mathcal{J}}R_{j}$ and $\vartheta \leftarrow 0$;
			\For{$l \in \mathcal{L}$}
			\If{$\sum_{j\in \mathcal{J}_{l}}u\left( R_{j}, \xi_{j}\right) <\eta_{l}$}
			\For{$j \in \mathcal{J}_{l}$}
			\If{$R_{j}<\xi_{j}$}
			\State $\vartheta\leftarrow \vartheta+(R_{j}-\xi_{j})/\xi_{j}$;
			\EndIf
			\EndFor
			\EndIf
			\EndFor
			\If{$\vartheta<0$}
			\State $\phi \leftarrow \vartheta /\phi$;
			\EndIf; \State \Return $\phi$;
		\end{algorithmic}
		\label{ALG:MET2}
	}
\end{algorithm}

\begin{algorithm}[t]
	{\scriptsize
		\caption{Deep $Q$-learning based Resource Assignment}
		\footnotesize
		\begin{algorithmic}[1]
			\State Initialize replay memory $\mathcal{D}$, $\gamma$ and $\tau$;
			\State Initialize the training DQN, $Q_{\mat{\theta}_{\text{train}}}$, with random weights $\mat{\theta}_{\text{train}}$;
			\State Initialize the target DQN, $Q_{\mat{\theta}_{\text{target}}}$, with weights $\mat{\theta}_{\text{target}}\leftarrow\mat{\theta}_{\text{train}}$;
			\Loop { over the episodes}
			\State Observe current state of all agents $s^{(n)}$;
			\State Each agent $n$ chooses an action $a^{(n)}$ using $\epsilon$-greedy policy from $Q_{\mat{\theta}_{\text{target}}}$;
			\State Execute the action of each agent, i.e., $\mathbf{a} \leftarrow [a^{(1)},\ \cdots, a^{(N)}]$;
			\State Obtain $\phi$ using Alg. \ref{ALG:MET2};
			\State Observe the next state of each agent ($s'^{(n)}$);
			\State Store experience $(s^{(n)},a^{(n)},\phi,s'^{(n)})$ of each agent in $\mathcal{D}$;
			\State Sample a set of random experiences, i.e., a mini-batch $\mathcal{B}$ from $\mathcal{D}$;
			\State Perform the gradient descent step on \eqref{L} with respect to the weights $\mat{\theta}_{\text{train}}$;
			\State At every $\tau$ episodes replaces target parameters, i.e., $\mat{\theta}_{\text{target}}\leftarrow\mat{\theta}_{\text{train}}$;
			\State Update the  $\epsilon$-greedy decision policy;
			\State $s^{(n)}\leftarrow s'^{(n)}, \quad \forall n \in \mathcal{N}$;
			\EndLoop;
			\State \Return $\mathbf{a}$;
		\end{algorithmic}
		\label{Alg:Proposed_Solution}
	}
\end{algorithm}

Our proposed solution based on deep $Q$-learning to problem \eqref{EQ:OBJ2} is shown in \AlgRef{Alg:Proposed_Solution}. Basically, the idea of this algorithm is to use the concepts of \ac{DQN} and those defined in Section \ref{SUBSEC:Overview_reinforcement-learning} to approximate $Q$-table by a function and, therefore, avoid the main disadvantages of the proposal addressed in \cite{Saraiva2019}, such as high memory cost.

Regarding \AlgRef{Alg:Proposed_Solution}, in lines 1, 2 and 3 we define the main structures for our approach. More specifically, in line 1, we reserve a limited amount of memory, $\mathcal{D}$, and define a certain number of iterations, $\tau$, that represent a period for updating target \ac{DQN} weights. In lines 2 and 3, we randomly initialize the target and training \acp{DQN} responsible for the learning process, where both \acp{DQN} are fully-connected \acp{DNN} that consists of four layers: an input layer, an output layer and two hidden layers in between. The input layer is fed by the state vector of length $(N+J)$ and the output layer has dimensionality corresponding to the number of possible actions, in this case, $J$.

\begin{figure}[t]
	\centering
	\includegraphics[width=\linewidth]{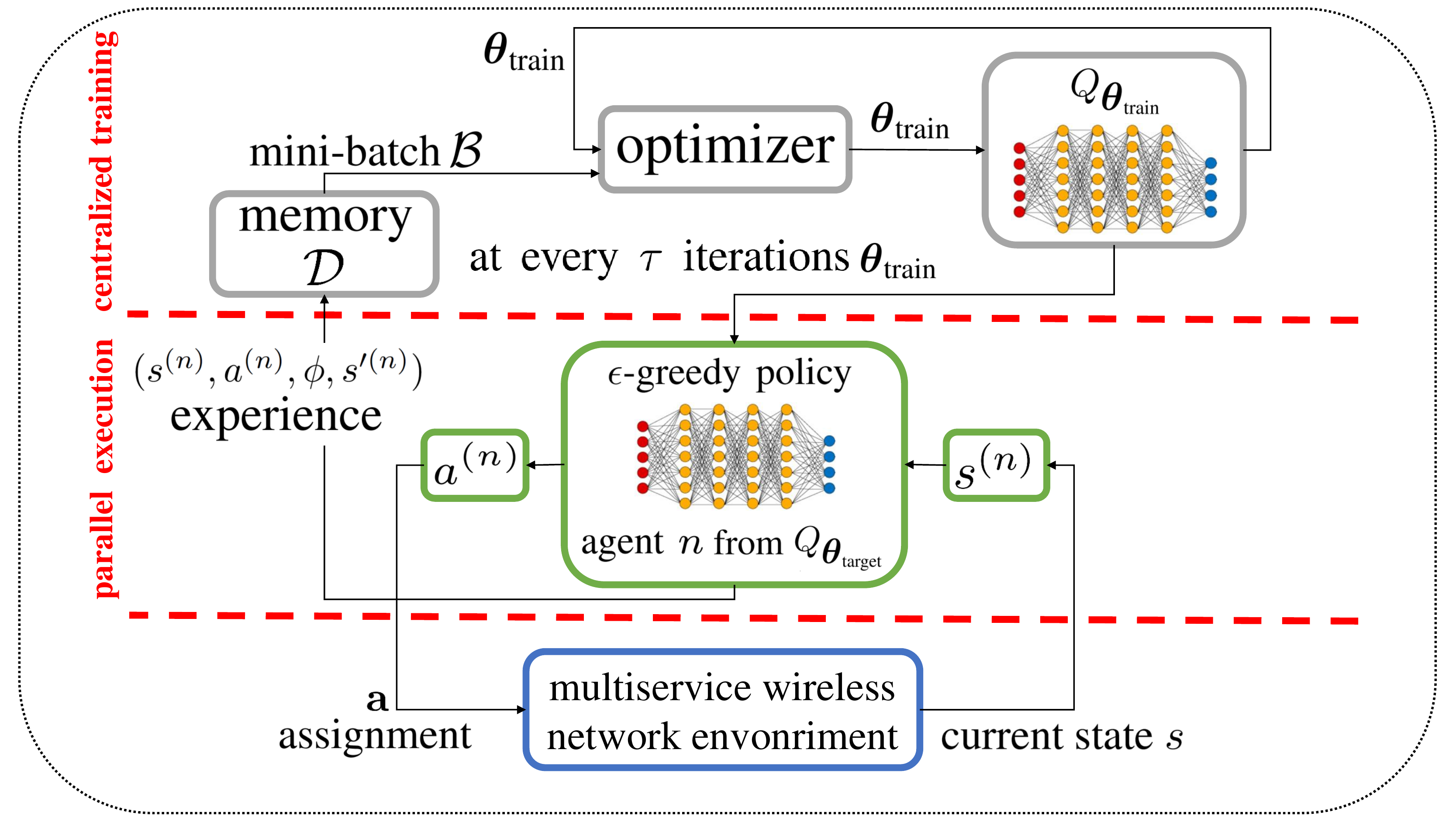}
	\caption{llustration of the proposed multi-agent deep \ac{RL} algorithm for resource allocation \cite{Nasir2019}, adapted.}
	\label{FIG:Scheme}
\end{figure}

The \textit{loop} between lines 4 and 16 represents the learning process, which is responsible for adjusting the weights of training and target \acp{DQN}, where each iteration is defined as an episode. We assume an approach in which the actions are taken in parallel by each agent while the training is performed by a central module according to \FigRef{FIG:Scheme}. In this figure, on one hand, it can be seen that the decisions of the $N$ agents can be chosen in parallel using a DQN whose input and output depend on the current states and possible actions of each agent, respectively. However, on the other hand, the training phase is centralized and, therefore, the experiences of all agents are constantly collected to adjust the weights of another DQN. This framework eases implementation and improves stability. Moreover, this strategy can also  significantly  reduce  the  amount of memory and computational resources required by training \cite{Nasir2019}. Therefore, each agent has the same copy of $Q_{\mat{\theta}_{\text{target}}}$, while $Q_{\mat{\theta}_{\text{train}}}$ is localized at the central module. Thus, in each episode, all agents observe their respective states and are synchronized to take their actions at the same time
based on $\epsilon$-greedy policy from $Q_{\mat{\theta}_{\text{target}}}$, according to lines 5 and 6. Next, an assignment pattern, $\mathbf{a}$, is defined from the agent's actions, the reward, $\phi$, is calculated and each agent $n$ observes its next state according to lines 7, 8 and 9, respectively.
Note that the reward is common for all agents so that they can benefit from each other's experiences to try to learn the optimal policy. In other words, the agents work collaboratively to maximize the obtained reward and, consequently, the objective in \eqref{EQ:Objective_function}.

After that, in line 10, we define an experience sample as a tuple, $(s^{(n)},a^{(n)},\phi, s'^{(n)})$, consisting in current
state $s^{(n)}$, chosen action $a^{(n)}$, reward $\phi$ and next state, $s'^{(n)}$ of each agent. In addition, in order to avoid oscillations and divergence in the parameters, we use the concept of memory replay so that the tuples of experiences of all agents are stored in memory $\mathcal{D}$. We consider that this memory is a \ac{FIFO} queue where a new experience replaces the oldest experience in the queue when the number of experiences
exceeds the capacity, $D$. %
In order to train the parameters $\mat{\theta}_{\text{train}}$, a mini-batch of experiences, $\mathcal{B}$, is sampled randomly from $\mathcal{D}$ and the stochastic gradient descent method is performed by central module to minimize the cost function in \eqref{L} as shown in lines 11 and 12, respectively. Furthermore, the process of updating the parameters $\mat{\theta}_{\text{target}}$ is periodic and, therefore, in line 13, only at every $\tau$ episodes, the new parameters $\mat{\theta}_{\text{train}}$ are available for target \ac{DQN}. Finally, in line 14, the  $\epsilon$-greedy policy is updated, the current state of each agent changes to the next state (line 15) and another episode starts.

Mathematically, the complexity of \AlgRef{Alg:Proposed_Solution} can be evaluated by quantifying the complexity to obtain the $Q$-function from the \ac{DQN} and to train the weights of the \ac{DQN} since this is the main idea of this algorithm. Obviously, it highly depends on the structure of the employed \ac{DQN} and its parameters. As discussed,  in our case, the \ac{DQN} is composed by fully-connected layers and, thereby, the complexity of the algorithm is given by $\mathcal{O}(wm \log m)$ where $w$ is the number of layers and $m$ is the number of units per layer \cite{Lee2019}.

\begin{figure}[t]
	\centering
	\includegraphics[width=\linewidth]{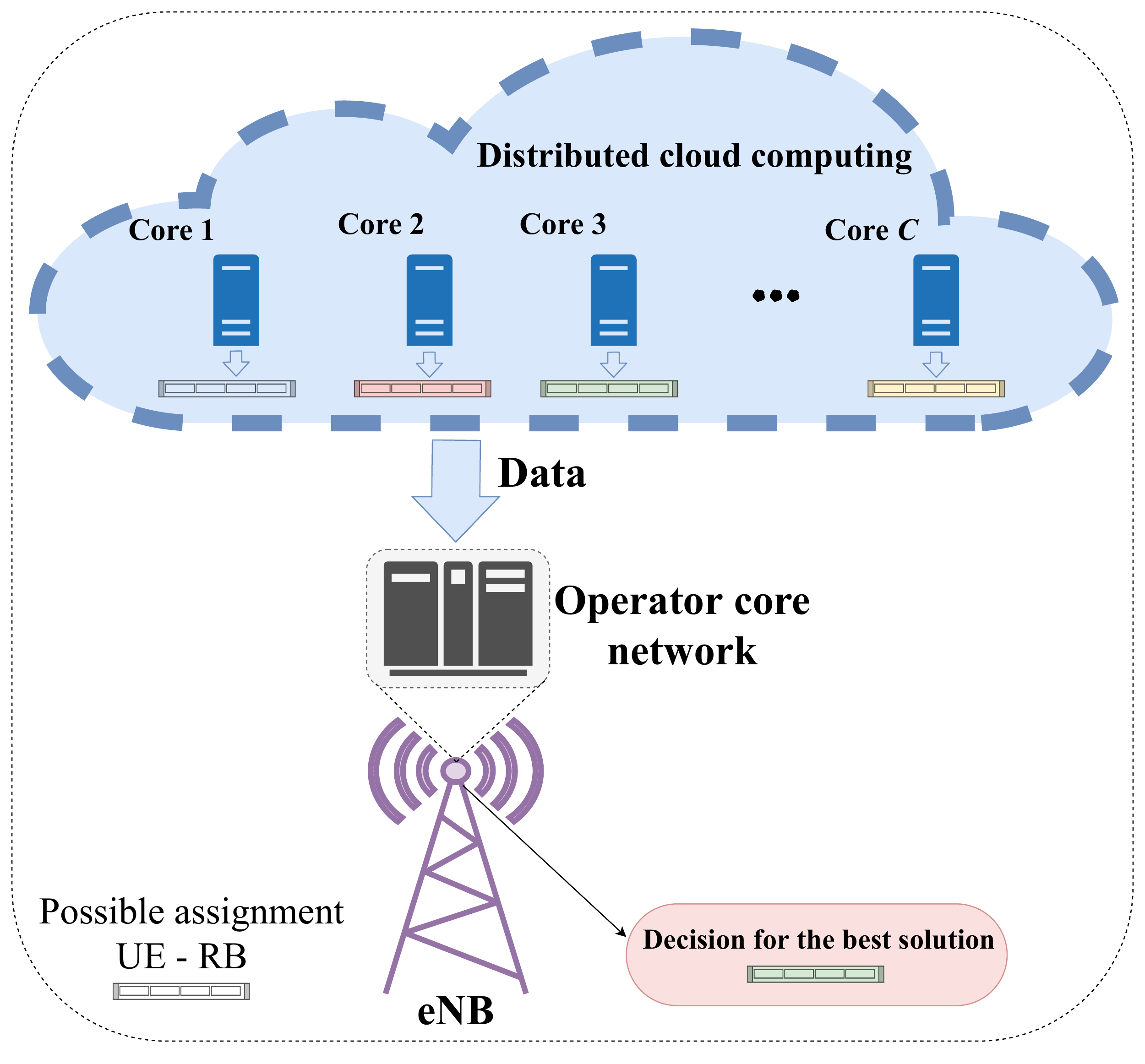}
	\caption{Proposed solution to problem \eqref{EQ:OBJ2} using parallel execution of \AlgRef{Alg:Proposed_Solution} on different cores.}
	\label{FIG:Scheme_Proposed_Solution}
\end{figure}

Something interesting about \AlgRef{Alg:Proposed_Solution} is that depending on the initialization of the \acp{DQN} weights or parameters, the algorithm can converge to a solution more or less accurately relative to the optimal solution of problem \eqref{EQ:OBJ2}, given a fixed number of episodes. Indeed, this can be exploited by letting \AlgRef{Alg:Proposed_Solution} run multiple times on different cores and, therefore, with totally independent weights initialization. As a result, since the runtime of each core is the same, the idea is to choose the best output as a solution to problem \eqref{EQ:OBJ2} as depicted in \FigRef{FIG:Scheme_Proposed_Solution}. Note that parallel execution of multiple cores does not necessarily need to be computed on the \ac{eNB} itself. Due to possible limitations of this infrastructure such as overhead, small storage space and low computing ability, the data storage and processing can be moved to decentralized and powerful computing platforms located in a cloud. In terms of performance, the cloud utilizes distributed system architectures and can offer excellent computation speeds. Besides, cloud computing provides many other advantages as quick deployment, easy integration, resiliency, redundancy, backup,  disaster recovery, among others \cite{Ahmed2014}. Thus, the \ac{eNB} is limited to deciding the best solution after data processing.
\section{Performance Evaluation} \label{SEC:PerfEval}

In this section, we evaluate our proposed solution and compare it with the optimal solution and with the solutions of \cite{Lima2012}, \cite{Sousa2016} and \cite{Saraiva2019}. %
We firstly present the main simulations parameters and, after that, the results and their discussion.

\subsection{Simulation Assumptions} \label{SEC:Sim_Assum}

We consider 6 \acp{RB} ($N=6$), 4 \acp{UE} ($J=4$), 2 service plans ($L=2$) and we admit that \acp{UE} from service plan $2$ demand a throughput of $150$ kbps higher than the \acp{UE} from the service plan $1$. %
In both services, we consider only two \acp{UE}, where $\eta_{1}=2$ and $\eta_{2}=1$. %
We assume $11$ \ac{QoS} levels in kbps such that $\xi_{j\in\mathcal{J}_1}=(150, 220, \dots, 850)$, i.e., the required data rates for service plan $1$ vary between $150$ kbps and $850$ kbps at the step of $70$ kbps. Consequently, the requeriments for service plan $2$ vary between $300$ and $1000$ with the same step. %
The DQN was implemented using Tensorflow \cite{Abadi2016}, assuming two hidden layers. We use the rectifier linear unit (ReLU) as DQN's
activation function and we use Adam's
algorithm \cite{Kingma2014} for the optimization. Moreover, we consider that
$\epsilon$-greedy policy varies over the episodes following an exponential decay. In general, all the important simulation parameters are shown in \TabRef{tab:dqn} and \TabRef{tab}. %

To perform qualitative comparisons with our proposed algorithm (deep $Q$-RA), we simulate the optimal solution of problem $\eqref{EQ:OBJ2}$ (OPT) as well as the algorithms Reallocation-based Assignment for
Improved Spectral Efficiency and Satisfaction (RAISES) \cite{Lima2012}, Rate Maximization under Experience Constraints (RMEC) \cite{Sousa2016} and $Q$-learning based Resource Assignment ($Q$-RA) \cite{Saraiva2019}. On one hand, RAISES and RMEC are traditional rule-based algorithms, which use resource reallocation strategies to define the best assignment pattern for the system. On the other hand, $Q$-RA, as its name suggests, is an algorithm based on $Q$-learning technique for resource allocation. Therefore, $Q$-RA algorithm is a tabular learning method, where a single agent accumulates all its experience in a $Q$-table over several episodes.

Regarding the performance metrics, we consider the outage rate and the system throughput. An outage event happens when an algorithm cannot manage to find a feasible
solution, i.e., the algorithm does not find a solution fulfilling the constraints of problem \eqref{EQ:OBJ2}. Then, outage rate is defined as the ratio between
the number of instances with outage events and the total number of simulated instances. The system throughput is the sum of the data rates obtained by all the \acp{UE} in a given instance. The results were obtained by running $1,000$ feasible instances of problem \eqref{EQ:OBJ2} in order to get valid results in a statistical sense and the channel realizations were the same for all the simulated algorithms to get fair comparisons.

\begin{table}[!t]
	\centering
	\caption{DQN Parameters}
		\begin{tabular}{>{\centering\arraybackslash}p{0.7\columnwidth}|>{\centering\arraybackslash}p{0.2\columnwidth}}
			\hline \hline
			\textbf{Parameter} & \textbf{Value}
			\\ \hline
			Memory size & 1000
			\\ \hline
			Mini-batch size & 256
			\\ \hline
			Number of neurons per hidden layer & 64
			\\ \hline
			Initial value for $\epsilon$ & 0.8
			\\ \hline
			Decay rate  & 0.001
			\\ \hline
			Learning rate  & 0.0001
			\\ \hline
			Discount factor & 0
			\\ \hline
			Period for updating
			target DQN weights & 5
			\\
			\hline \hline
	\end{tabular}%
	\label{tab:dqn}
\end{table}
\begin{table}[!t]
	\centering
	\caption{Network Parameters \cite{Lima2012}}
		\begin{tabular}{>{\centering\arraybackslash}p{0.5\columnwidth}|>{\centering\arraybackslash}p{0.4\columnwidth}}
			\hline \hline
			\textbf{Parameter} & \textbf{Value}
			\\ \hline
			Cell radius & \SI{334}{\m}
			\\ \hline
			Transmit power per \ac{RB} & \SI{0.35}{\watt}
			\\ \hline
			Number of subcarriers per \ac{RB} & 12
			\\ \hline
			Shadowing standard deviation& \SI{8}{\decibel}
			\\ \hline
			Path loss & 35.3 + 37.6$\ \cdot\log(d)$ [\SI{}{\decibel}]
			\\ \hline
			Traffic model & Full buffer
			\\ \hline
			Noise spectral density & 3.16$\ \cdot 10^{-20}$ \SI{}{\watt}/\SI{}{\Hz}
			\\
			\hline \hline
	\end{tabular}%
	\label{tab}
\end{table}

\begin{figure}[b]
	\centering
	\includegraphics[width=\linewidth]{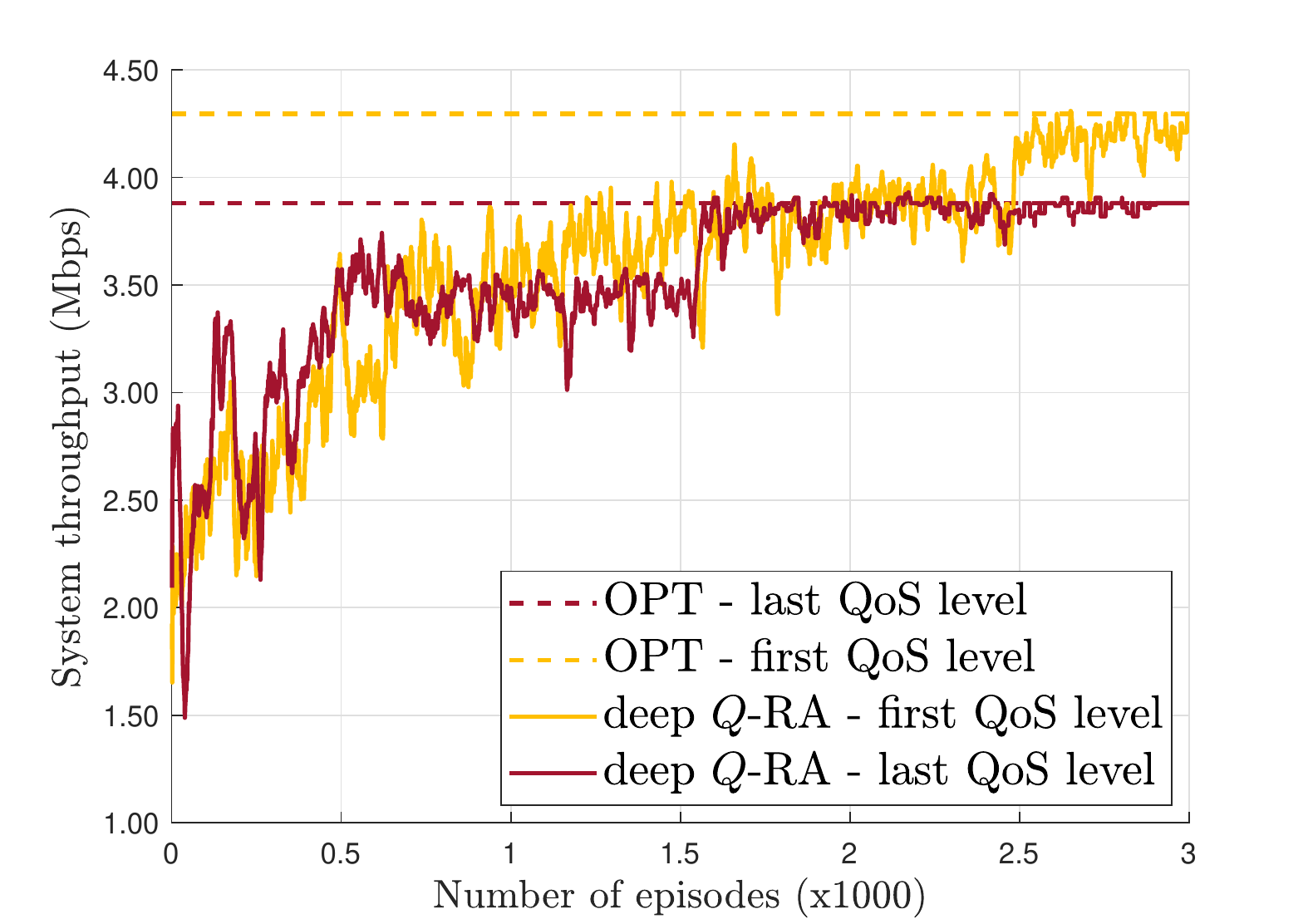}
	\caption{System throughput versus number of episodes in a particular instance for OPT and deep $Q$-RA algorithms.}
	\label{RES0}
\end{figure}
\subsection{Numerical Results}

\FigRef{RES0} shows the system throughput versus the number of episodes for the algorithms OPT and deep $Q$-RA, condesidering the first and the last \ac{QoS} levels described in Section \ref{SEC:Sim_Assum}. Looking at the performance of the deep $Q$-RA solution, we can observe that it converges to the OPT solution as the number of episodes increases for both investigated QoS levels. This is an expected result since the more episodes we have, the more accurate the estimation of $Q$-function is and, consequently, the more favorable it is for the agents to converge to the optimal solution of problem \eqref{EQ:OBJ2}. Moreover, note that in \FigRef{RES0} the convergence time to the optimal solution may vary depending on the required QoS level. This is because at low QoS levels there are several possible solutions and, as a result, it can be more difficult to converge to the optimal solution of problem \eqref{EQ:OBJ2}. For scenarios with high QoS levels required, possible solutions are rarer but once found means near optimal solutions, consequently the deep $Q$-RA algorithm tends to focus on them. Indeed, this can lead to faster convergence.

\begin{figure}[t]
	\centering
	\includegraphics[width=\linewidth]{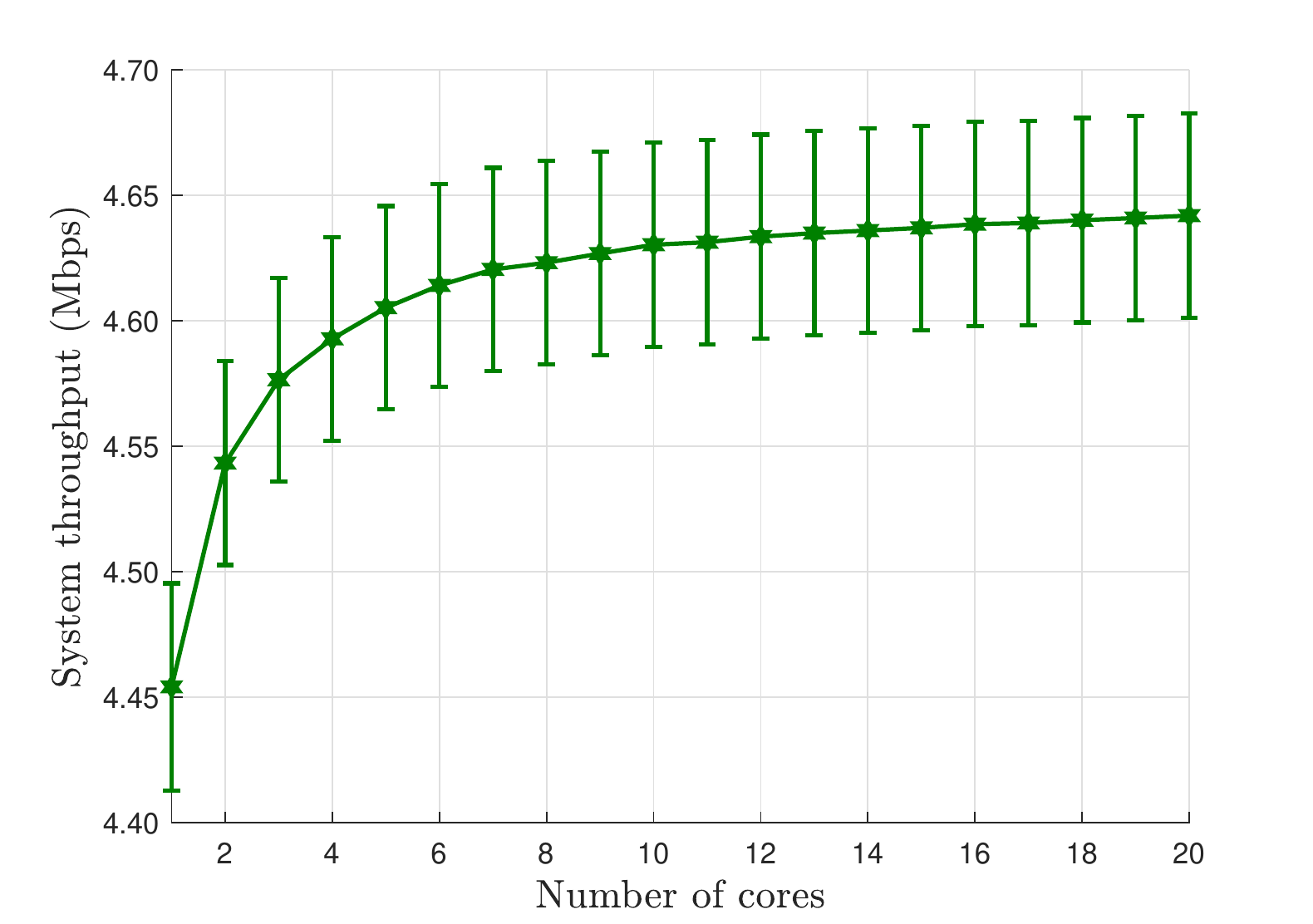}
	\caption{System throughput of deep $Q$-RA algorithm versus the number of parallel cores in the system.}
	\label{RES1}
\end{figure}

\begin{figure}[t]
	\centering
	\includegraphics[width=\linewidth]{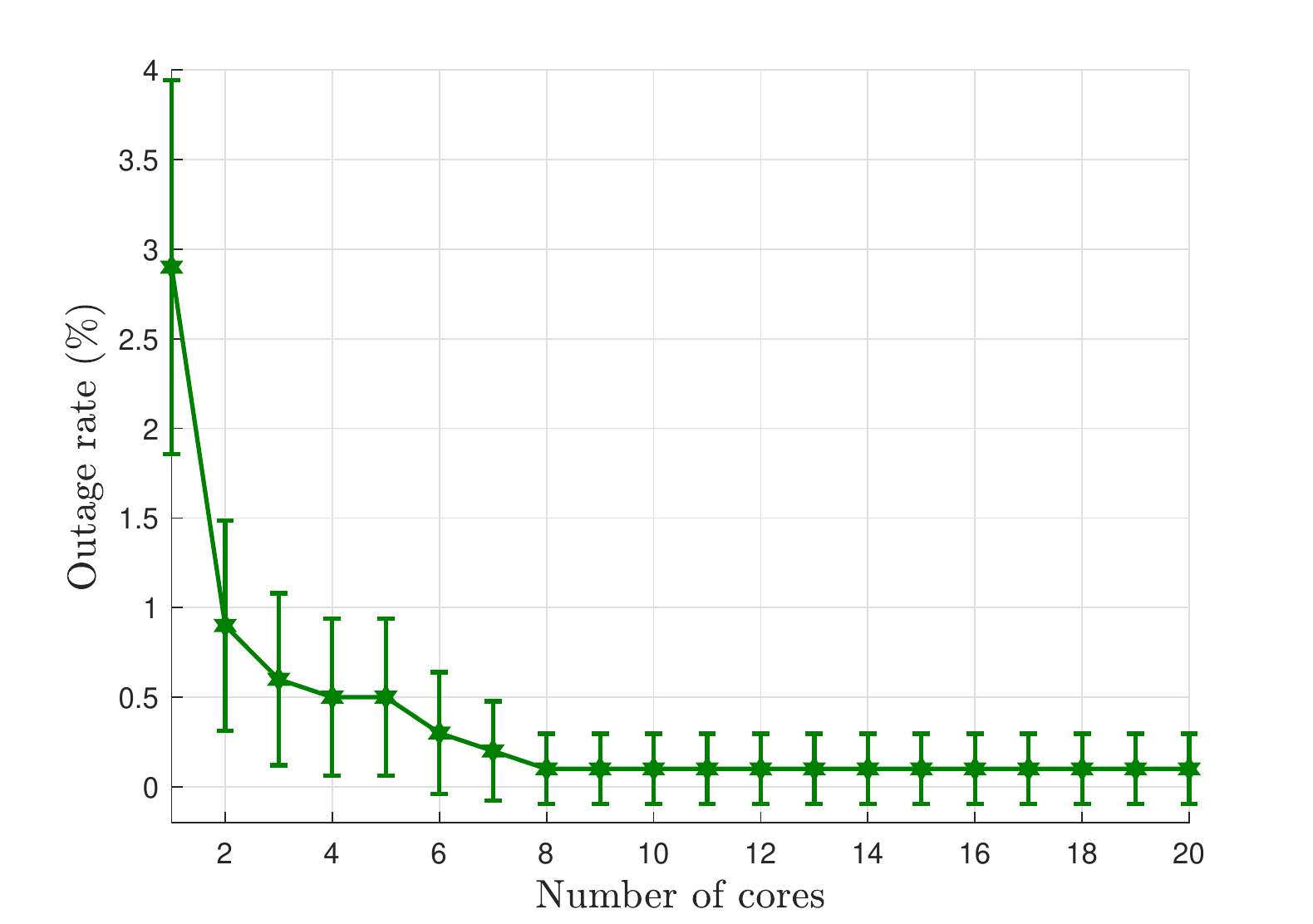}
	\caption{Outage rate of deep $Q$-RA algorithm versus the number of parallel cores in the system.}
	\label{RES2}
\end{figure}

In \FigRef{RES1} and \FigRef{RES2}, we plot the system throughput and outage rate versus the number of parallel cores in the system, respectively, in order to show the advantages of the structure illustrated in \FigRef{FIG:Scheme_Proposed_Solution}. Also, from here, we assume for all the following results a confidence interval with a $95\%$ confidence level. Firsty, in \FigRef{RES1} and \FigRef{RES2}, note that as the number of cores in the system increases, there is a considerable increase in the performance of the proposed solution. In addition, due to the characteristics of the deep $Q$-learning technique, this structure does not require a high memory consumption and, as shown in the last figures, a relatively low number of cores is enough to ensure excellent performance. Note, for example, that with less than $10$ cores there is practically no outage in the system for the investigated scenario.

\begin{figure}[t]
	\centering
	\includegraphics[width=\linewidth]{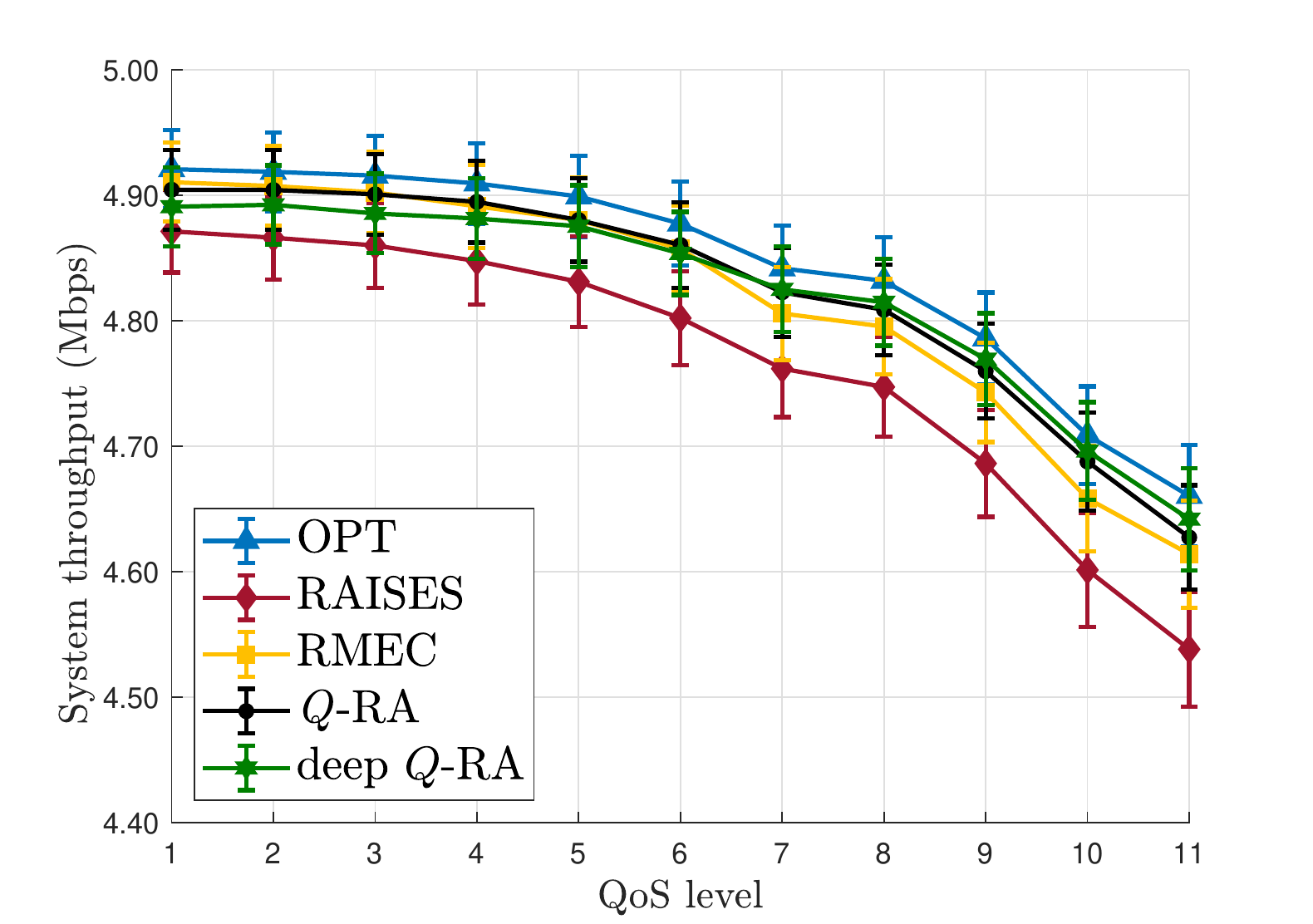}
	\caption{System throughput versus QoS level for OPT, RAISES, RMEC, $Q$-RA and deep $Q$-RA algorithms in the considered scenario.}
	\label{RES3}
\end{figure}
\begin{figure}[t]
	\centering
	\includegraphics[width=\linewidth]{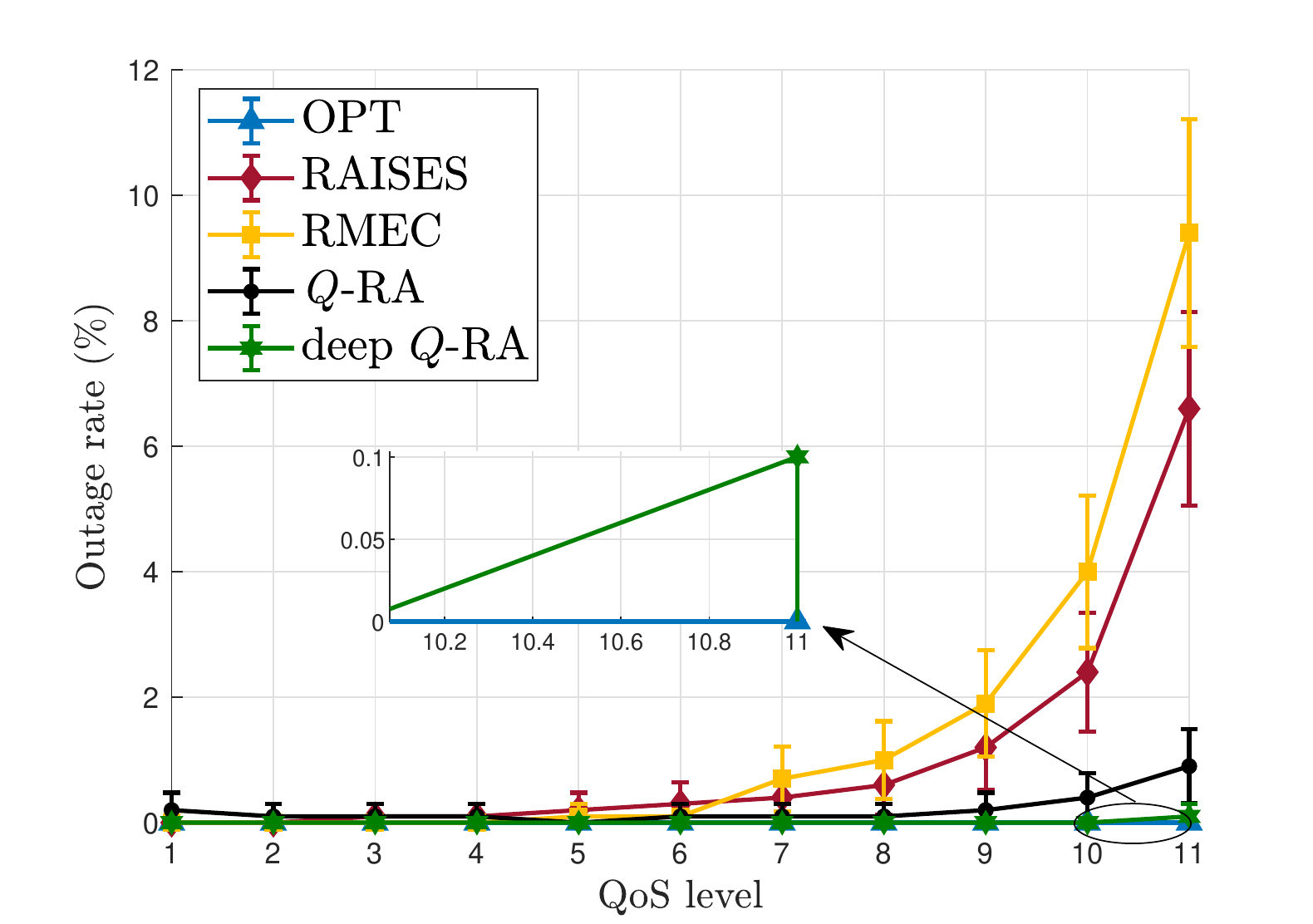}
	\caption{Outage rate versus QoS level for OPT, RAISES, RMEC, $Q$-RA and deep $Q$-RA algorithms in the considered scenario.}
	\label{RES4}
\end{figure}

Now we compare our approach with other proposals from the literature. In \FigRef{RES3} and \FigRef{RES4}, we plot the system throughput and the outage rate in the considered scenario versus the \ac{QoS} level for the algorithms OPT, RAISES, RMEC, $Q$-RA and deep $Q$-RA, respectively. %
For the $Q$-RA and deep $Q$-RA algorithms, we consider $3,000$ episodes in the plots of these figures. Besides, for deep $Q$-RA algorithm we use $10$ cores. %
In this way, we firstly observe a near optimal performance of our proposed solution both in terms of outage rate and system throughput to problem \eqref{EQ:OBJ2}. %
In fact, we highlight \FigRef{RES4} that shows the outage curve that is considerably better for the solutions based on \ac{RL}, with even better performance for deep $Q$-RA solution. %
In this figure, notice that the outage rate for these solutions are smaller than $1\%$ and $0.1\%$ for $Q$-RA and deep $Q$-RA, respectively.

On the other hand,  RAISES and RMEC solutions have much higher outage rates, with approximately $7\%$ and $10\%$ for the highest \ac{QoS} level, respectively. %
This shows that solutions based on \ac{ML} algorithms may perform better than traditional heuristics and, therefore, they can be considered as a promising tool to solve resource allocation problems in modern networks. However, as highlighted in \cite{Saraiva2019}, $Q$-RA solution may require a high memory cost to build and store $Q$-table because it directly depends on space $\mathcal{S}\times \mathcal{A}$. Therefore, this makes its use more difficult in interesting and realistic scenarios. As discussed earlier, this is not a problem for deep $Q$-RA solution, which may in fact become a more attractive and less problematic solution in larger scenarios.

\section{Conclusions and Perspectives} \label{SEC:Conclusion}
In this paper, we have investigated the problem of maximizing the system throughput subject to user satisfaction ratio constraints in a multiservice scenario. %
This problem was previously studied in \cite{Lima2012}, \cite{Sousa2016} and \cite{Saraiva2019}, where tradicional heuristics or machine learning based methods were proposed. %
However, to tackle this problem we have proposed a new decentralized radio resource allocation mechanism employing multi-agent deep reinforcement learning. %
From the simulation results, we have shown that each agent can learn how to jointly deal with resource allocation and QoS garantees while maximizing the system throughput. As a result, our proposed can provide better performance than the other benchmark approaches simulated in this article.

Regarding future works, we believe that the proposed framework in this paper can be improved by taking into consideration the channel correlation along the time and redefining the system state in order to considerably decrease the need for training when applied in dynamic contexts.
Finally, other approaches where learning-based techniques are jointly responsible for allocating power and resource can also be analyzed in the future.%

\ifCLASSOPTIONcaptionsoff
  \newpage
\fi

\printbibliography
\end{document}